\documentclass[conference]{IEEEtran}
\usepackage[utf8]{inputenc}
\usepackage{amsmath,amssymb,amsfonts}
\usepackage{algorithmic}
\usepackage{graphicx}
\usepackage{textcomp}
\usepackage{xcolor}
\usepackage{fancyhdr}
\usepackage[hyphens]{url}

\def\BibTeX{{\rm B\kern-.05em{\sc i\kern-.025em b}\kern-.08em
    T\kern-.1667em\lower.7ex\hbox{E}\kern-.125emX}}

\title{SoftWear: Software-Only In-Memory Wear-Leveling \\for
Non-Volatile Main Memory} 
\author{Christian Hakert$^\ast$, Kuan-Hsun Chen$^\ast$, Paul R. Genssler$^\dag$, Georg von der Brüggen$^\ast$, Lars Bauer$^\dag$, Hussam Amrouch$^\dag$,\\Jian-Jia Chen$^\ast$, Jörg Henkel$^\dag$\\
\textit{$^\ast$ Design Automation for Embedded Systems Group, TU Dortmund Unibersity, Germany}\\
\textit{$^\dag$ Chair for Embedded Systems, KIT, Germany}}

\usepackage[nocompress]{cite}

\usepackage{tikz}
\usepackage[europeanresistors,americaninductors]{circuitikz}
\usetikzlibrary{positioning,chains,arrows,shadows,fadings,shapes,backgrounds,snakes,matrix,patterns,plotmarks,trees,mindmap, calc}

\definecolor{TUgreen}{HTML}{649600}

\newcommand{\revised}[1]{{#1}}

\usepackage{cleveref}
\usepackage{xspace}
\newcommand{\eg}{e.g.\@\xspace}
\newcommand{\ie}{i.e.\@\xspace}
\newcommand{\etal}{et al.\@\xspace}

\newcommand{\logplotheight}{.28pt}

\begin{document}
\maketitle
\thispagestyle{plain}
\pagestyle{plain}


\begin{abstract}
Several emerging 
technologies for byte-addressable non-volatile memory (NVM) have
been considered to replace 
DRAM as the main memory in computer systems
during the last years. The disadvantage of a lower write endurance, compared to
DRAM, of NVM technologies like Phase-Change Memory (PCM) or Ferroelectric RAM
(FeRAM) has been 
addressed in the literature. 
As a solution, in-memory wear-leveling techniques have been proposed, which
aim to balance the wear-level over all memory cells to achieve an increased memory 
lifetime. 
Generally, to apply such advanced aging-aware wear-leveling techniques proposed 
in the literature, additional special hardware is introduced into the
memory system to 
provide the necessary information about the cell age 
and thus 
enable aging-aware wear-leveling decisions.


This paper proposes software-only aging-aware wear-leveling based on common CPU
features and does not rely on any additional hardware support from the
memory subsystem. Specifically, we exploit the memory management unit (MMU),
performance counters, and interrupts to approximate the
memory write counts as an aging indicator. Although the software-only approach may lead to  
slightly worse wear-leveling, it is applicable on commonly available hardware.
We achieve
page-level coarse-grained wear-leveling by approximating the current cell
age through statistical sampling and performing physical memory
remapping through the MMU.
This method results in non-uniform memory usage patterns within a
memory page. Hence, we further propose a fine-grained wear-leveling in the stack region of C / C++ compiled
software. 

By applying both wear-leveling techniques, we achieve up to $78.43\%$ of the
ideal memory lifetime, which  is a lifetime improvement of more than a factor of
$900$ compared to the lifetime without any wear-leveling.

\end{abstract}

\section{Introduction}
Emerging technologies for non-volatile memory (NVM), like Phase-Change-Memory
(PCM) or Ferroelectric RAM (FeRAM), have been considered as a replacement for DRAM as the main memory over
the last years. Most NVM technologies feature advantages like low energy
consumption and high integration density, which makes them a desired main memory
replacement. One of the major disadvantages of some NVM technologies is the
lower write-endurance. While classic DRAM endures for more than $10^{15}$ write
cycles, PCM only endures $10^8-10^9$ write cycles per cell \cite{nvmsurvey}.
Thus, to wear out a DRAM cell within 10 years, an application would have to
write the same memory cell every $900^{th}$ CPU cycle in average on a 3GHz CPU.
Applying the same application to PCM, the memory would wear-out within 5
minutes. Although typical applications do not cause such an extreme write
pattern, they still cause a highly non-uniform write pattern to the memory
\cite{Ferreira:2010:IPM:1870926.1871147, startgap, 5375306}. Accordingly, the
problem has been tackled in the literature and several in-memory wear-leveling
techniques have been proposed. A majority of these techniques is aging-aware \revised{\cite{stackalloc,
dong2011wear, 6509609, 7046386, AghaeiKhouzani:2014:PPL:2627369.2627667,
7410326, Chen:2012:APW:2228360.2228439, Ferreira:2010:IPM:1870926.1871147,
Zhou:2009:DEE:1555754.1555759, 227782}},
which means that the current cell age or the current write count is taken into account for the
wear-leveling decisions. The wear-leveling itself is mostly realized through an
abstraction layer, which remaps the physical location of logical memory regions.
However, as current memory hardware does not provide a write-count, which is
necessary to determine the cell age, additional hardware is introduced. This
hardware requires additional chip-space, and might be hard to realize in a way
that meets the desired granularity and clock-frequency.

To allow aging-aware wear-leveling in the absence of such special hardware,
this paper proposes software-only wear-leveling techniques. The term
software-only here means that we do not require any additional hardware from the
memory subsystem and only use hardware features which are widely available. 
We provide the necessary write-count through a 
statistical online approximation of the write distribution, which only requires
a memory management unit (MMU), performance counters, and an interrupt mechanism.
The performance counter allows to generate an interrupt every $n^{th}$ memory
write access, which achieves an equidistant sampling of the write distribution.
A special configuration of the memory access permission allows to record the
target address of a single memory write afterwards. The approximated write
distribution 
enables an arbitrary aging-aware wear-leveling algorithm
subsequently. In this paper, we implement a simple wear-leveling algorithm on
the granularity of virtual memory pages, which achieves the necessary physical
memory remapping through the MMU. 
Since  the
resulting memory write distribution still 
results in high non-uniformity due to the granularity of memory pages, we
introduce an additional software-only, fine-grained wear-leveling technique,
which balances the write-accesses to the stack region by relocating the stack in
a circular manner. This is achieved by copying the current stack content
regularly to a new location and adjust the stack-pointer accordingly. A special
virtual memory configuration allows a hardware-aided wraparound to achieve a
circular movement.
\newpage
\noindent\textbf{Our contributions:}
\begin{itemize}
 \item We deliver a software-only coarse-grained in-memory wear-leveling system,
 consisting of an online approximation mechanism for the write-distribution and an MMU-based wear-leveling algorithm.
 \item We further provide an extending software-only fine-grained wear-leveling 
 technique, which targets the stack region of C/C++ compiled applications and relocates the stack in a circular manner in a bounded memory region.
\end{itemize}

We aim to balance the write-count to each memory byte in the 
flat memory space equally to achieve a high memory lifetime.
We note that other factors impact the memory endurance as well, \eg, process
variation in PCM~\cite{5375306}, but the write-count is a major factor.  
Our approaches can be
extended according to physical models (\eg process variation domains) to also 
respect advanced physical memory properties.

After giving an overview about the related wear-leveling approaches in
literature in \Cref{sec_relatedwork}, we present the memory write distribution
of our benchmark applications in \Cref{sec_appmem} and our method to analyze the
write pattern of applications, which is also used for our evaluations in
\Cref{sec_memanalysis}. After this, our novel wear-leveling techniques are
described in detail in \Cref{sec_coarsegrained} and \Cref{sec_finegrained}. Each
section contains an evaluation, which uses the write pattern analysis mechanism.
The paper concludes with a short summary in \Cref{sec_conclusion}.

\section{Related Work}
\label{sec_relatedwork}
During the last years, several approaches for in-memory wear-leveling for NVM
have been proposed. 
These approaches can be categorized along
different criteria. First, there are aging-aware approaches \cite{stackalloc,
dong2011wear, 6509609, 7046386, AghaeiKhouzani:2014:PPL:2627369.2627667,
7410326, Chen:2012:APW:2228360.2228439, Ferreira:2010:IPM:1870926.1871147,
Zhou:2009:DEE:1555754.1555759, 227782}, which take the current cell age into account to
apply wear-leveling. In contrast there are random-based approaches
\cite{startgap, Ferreira:2010:IPM:1870926.1871147,
Zhou:2009:DEE:1555754.1555759}, which apply wear-leveling in a circular
or random-based manner. 
Both approaches are 
often combined to 
achieve a random-based wear-leveling on fine granularities inside memory
blocks, while 
an aging-aware approach is used to target these 
coarse-grained memory blocks. The granularity also varies from single bits
\cite{Cho:2009:FSD:1669112.1669157, 6974656} over cache-lines \cite{startgap,
Zhou:2009:DEE:1555754.1555759} for fine-grained approaches to memory pages
\cite{Ferreira:2010:IPM:1870926.1871147,
AghaeiKhouzani:2014:PPL:2627369.2627667, 7410326, Chen:2012:APW:2228360.2228439, 227782}
or even bigger memory segments \cite{Zhou:2009:DEE:1555754.1555759, 5375306} for coarse-grained approaches.

Some approaches are not based on remapping the physical memory content through
an abstraction layer, but hook into the memory allocation process of the
operating system to apply wear-leveling to the memory allocator \cite{7410326,
AghaeiKhouzani:2014:PPL:2627369.2627667, stackalloc}. Li \etal~\cite{stackalloc}
also propose to use an allocated memory portion whenever a function is called for the function's
stack memory  to wear-level the stack region.

\revised{Gogte \etal propose a software-only coarse-grained wear-leveling approach by using a sampled approximation of the write distribution \cite{227782}. They make use of advanced debugging capabilities, \eg Intel Processor Event Based Sampling (PEBS), which allows them to sample the write requests from the CPU. These debugging capabilities, however, can rarely be found in embedded systems and resource constrained hardware.}

All \revised{other} mentioned aging-aware approaches rely on the the current write-count information
of the memory. Most approaches introduce specialized hardware into the memory
controller to collect the write-count information, which is not available in commonly
available systems and might be hard to realize. Dong \etal~\cite{dong2011wear}
use an offline recorded memory trace to estimate the write distribution,
which limits the approach to a subset of well-known applications only.

\section{Problem Description}
\label{sec_appmem}
When considering non-volatile memory as the main memory for program executions,
the system may suffer from the low write-endurance of the underlying memory
technology. Even if the system is also equipped with DRAM, certain applications
may be desired to only run on the non-volatile memory to reach energy saving
states as fast as possible. To understand the impact of program executions on
main memory with low write-endurance, the precise write distribution from a
program should be recorded and analyzed. Separating the program's memory
into the \texttt{text}, \texttt{data}, \texttt{bss}, and \texttt{stack} regions
allows to analyze the write pattern of each region separately and 
determine the impact on the memory lifetime. This section presents the write
distribution for our benchmark applications and points out the influence on the
write-endurance.

To determine the influence of code executions on the memory write-patterns of applications, especially on the different memory regions, we run four benchmark
applications.
We aggregate the resulting memory trace file on the granularity of 64 byte (a
cache-line is assumed to be written always entirely) to a write-count
distribution and present them graphically. As the benchmark applications we
chose following programs:
\begin{enumerate}
 \item \textbf{bitcount}: A simple implementation, which iterates over an array 
 of data and counts the 1 bits. The resulting count is stored in global counter and returned at the end.
 \item \textbf{pfor}: A simulation of a data decompression scenario. A big set
 of  data is available in a lightweight compressed format, namely
 \textit{Patched Frame of Reference}  (PFOR) \cite{zukowski2006super}. The data 
 is decompressed and aggregated in fixed size windows, which simulates the 
 processing of a stream of compressed data.
 \item \textbf{sha}: This application is part of the MiBench security suite 
 \cite{Guthaus:2001:MFC:1128020.1128563} and calculates the sha sum of a given 
 dataset.
 \item \textbf{dijkstra} This application is also part of the MiBench network 
 suite \cite{Guthaus:2001:MFC:1128020.1128563} and calculates a fixed number  of
 shortest paths in a network, using the dijkstra algorithm.
\end{enumerate}
We chose these benchmarks, because they are simple enough to understand the
connection  between the code and the memory usage of the different segments. The
limitation to four benchmarks is due to the high time consumption of the required full system simulations.

\begin{figure}
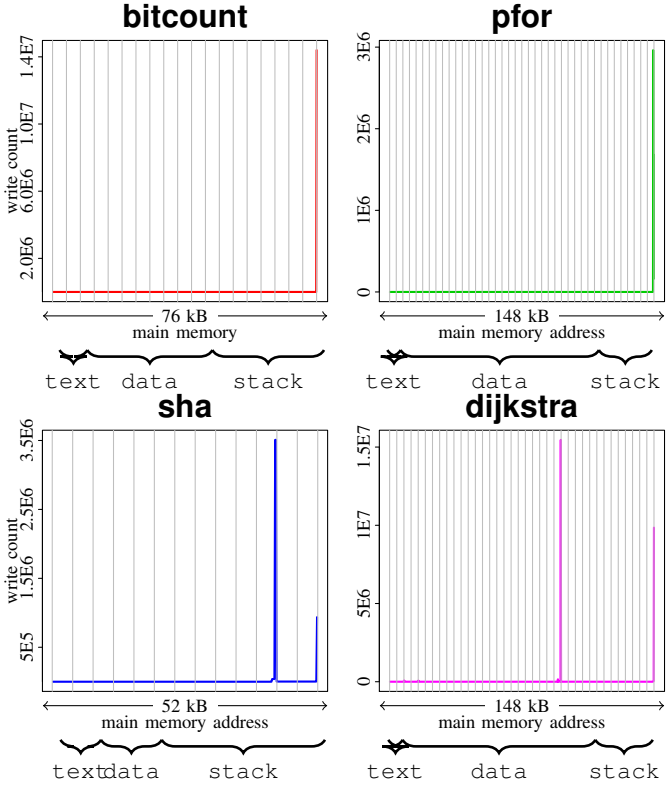

	\begin{center}
		 \begin{minipage}{.49\textwidth}
			\begin{minipage}{.49\textwidth}
				\input{figures/bitcount_baseline.tikz}
			\end{minipage}
			\begin{minipage}{.49\textwidth}
				\input{figures/pfor_baseline.tikz}
			\end{minipage}
		\end{minipage}
		
	\begin{tikzpicture}
	\newcommand{\scalefac}{1}
	\node[draw=white, rectangle, minimum width=1pt, minimum height=1pt] () at (-0.45cm,0) {};
	
	\draw [decoration={brace, amplitude=5pt}, decorate, line width=1pt] (10.5pt/\scalefac,0pt) -- node[below, yshift=-5.5pt] {\texttt{\small{text}}} (0pt/\scalefac,0pt);
	
	\draw [decoration={brace, amplitude=5pt}, decorate, line width=1pt] (57.8pt/\scalefac,0pt) -- node[below, yshift=-5pt] {\texttt{\small{data}}} (10.5pt/\scalefac,0pt);
	
	\draw [decoration={brace, amplitude=5pt}, decorate, line width=1pt] (100pt/\scalefac,0pt) -- node[below, yshift=-5pt] {\texttt{\small{stack}}} (57.8pt/\scalefac,0pt);
	
	\newcommand{\gap}{124pt}
	
	\draw [decoration={brace, amplitude=5pt}, decorate, line width=1pt] (5pt/\scalefac+\gap,0pt) -- node[below, yshift=-5.5pt] {\texttt{\small{text}}} (0pt/\scalefac+\gap,0pt);
	
	\draw [decoration={brace, amplitude=5pt}, decorate, line width=1pt] (80pt/\scalefac+\gap,0pt) -- node[below, yshift=-5pt] {\texttt{\small{data}}} (5pt/\scalefac+\gap,0pt);
	
	\draw [decoration={brace, amplitude=5pt}, decorate, line width=1pt] (100pt/\scalefac+\gap,0pt) -- node[below, yshift=-5pt] {\texttt{\small{stack}}} (80pt/\scalefac+\gap,0pt);
	\end{tikzpicture}
	
	\begin{minipage}{.49\textwidth}
			\begin{minipage}{.49\textwidth}
				\input{figures/sha_baseline.tikz}
			\end{minipage}
			\begin{minipage}{.49\textwidth}
				\input{figures/dijkstra_baseline.tikz}
			\end{minipage}
	\end{minipage}
	\begin{tikzpicture}
	\newcommand{\scalefac}{1}
	\node[draw=white, rectangle, minimum width=1pt, minimum height=1pt] () at (-0.45cm,0) {};
	
	\draw [decoration={brace, amplitude=5pt}, decorate, line width=1pt] (15.3pt/\scalefac,0pt) -- node[below, yshift=-5.5pt] {\texttt{\small{text}}} (0pt/\scalefac,0pt);
	
	\draw [decoration={brace, amplitude=5pt}, decorate, line width=1pt] (38.3pt/\scalefac,0pt) -- node[below, yshift=-5pt] {\texttt{\small{data}}} (15.3pt/\scalefac,0pt);
	
	\draw [decoration={brace, amplitude=5pt}, decorate, line width=1pt] (100pt/\scalefac,0pt) -- node[below, yshift=-5pt] {\texttt{\small{stack}}} (38.3pt/\scalefac,0pt);
	
	\newcommand{\gap}{124pt}
	
	\draw [decoration={brace, amplitude=5pt}, decorate, line width=1pt] (5.4pt/\scalefac+\gap,0pt) -- node[below, yshift=-5.5pt] {\texttt{\small{text}}} (0pt/\scalefac+\gap,0pt);
	
	\draw [decoration={brace, amplitude=5pt}, decorate, line width=1pt] (78.3pt/\scalefac+\gap,0pt) -- node[below, yshift=-5pt] {\texttt{\small{data}}} (5.4pt/\scalefac+\gap,0pt);
	
	\draw [decoration={brace, amplitude=5pt}, decorate, line width=1pt] (100pt/\scalefac+\gap,0pt) -- node[below, yshift=-5pt] {\texttt{\small{stack}}} (78.3pt/\scalefac+\gap,0pt);
	\end{tikzpicture}
		
	\end{center}
	\vspace{-0.5cm}
	\caption{Memory write-count distribution - baseline\protect\footnotemark}
	\vspace{-0.5cm}
	\label{memoryanalysis_baseline_dist}
\end{figure}

\Cref{memoryanalysis_baseline_dist} shows the resulting illustration of the
write-count distributions of the benchmark applications. Note that the four applications face different execution times and thus the total amount of writes is different. Thus, the scaling of the y axes is different. Considering the
different memory regions, different observations can be made:
\begin{itemize}
    \item \texttt{text}: As the \texttt{text} segment only contains the compiled
    binary code, it is never written during the normal application execution.
    This behavior 
    is also shown in the result. In the context of
    wear-leveling, read only memory regions have to be targeted as well as heavy
    written memory regions to distribute the wear-levels equally.
    \item \texttt{data/bss}: The \texttt{data} and the \texttt{bss} segments
    store global program variables, such as global attributes or
    arrays. Naturally, these variables are written from time to time, depending
    on the application logic. 
    The \textbf{dijkstra} benchmark has a heavy,
    non-uniform usage of the \texttt{bss} segment, 
    since the
    benchmark 
    manages the steps of the algorithm in a queue.
    \item \texttt{stack}: 
    The \texttt{stack} segment causes the most non-uniform write access to the main
    memory. This results 
    from the way the stack is typically used:
    Local variables are stored on  top of the stack and are removed when they
    are no longer used. Depending on the application logic, this makes the beginning of
    the stack a heavily used area with a lot of memory writes, while the rest of
    the stack region is used less. A wear-leveling algorithm has to distribute the
    memory writes to this region to all other, less written memory regions.
\end{itemize}
These results point out the need for aging-aware wear leveling. The memory
writes to  hot memory regions have to be redirected mainly to unused memory
regions,  but also to less used memory regions. This requires a monitoring of
the  current write-count and an incremental redistribution according to the current write-count distribution.

\footnotetext{The gray lines indicate boundaries of 4 kB virtual memory pages. The \texttt{data} and \texttt{bss} segment is marked as a big \texttt{data} segment	in the picture.}

\section{Memory Write-Pattern Analysis}
\label{sec_memanalysis}
\Cref{sec_appmem} presents the memory write-count distribution of four benchmark
applications.
In a usual 
computation platform, the memory accesses of a program cannot
be captured and analyzed without special techniques. 
Debugging mechanisms
can overcome the problem but introduce a 
large overhead. Using a hardware
analyzer, which basically plugs an FPGA between the CPU and the memory DIMM, is
considered by Bao~\etal~\cite{Bao:2008:HPI:1375457.1375484}. Such an analyzer
is reasonably fast but requires a complex hardware setup. 
In this paper,  we use a full system cycle-accurate simulator 
(including CPU, memory, buses, peripherals, etc.) on top of a Linux
host instead. This section introduces our simulation environment, which is also
used for the results in \Cref{sec_appmem}.

We chose 
\emph{gem5} 
\cite{Binkert:2011:GS:2024716.2024718} as the
full system simulator, since it can be combined with a memory
simulator for non-volatile memories, namely \emph{NVMain2.0} \cite{7038174}, due
to its modular structure. This setup allows  to obtain 
all memory accesses of a running program in a logfile, analyze them
afterwards,
and perform detailed evaluations of our methods by comparing the captured
logfiles.
To simulate the properties of NVMs, several simulators can be considered (\eg
\cite{volos2015quartz} and \cite{dong2012nvsim}), which precisely simulate,  for
instance, the timing and energy behavior. 
However, the methods in this paper
analyze and change the write behavior of applications only, which is
independent from the physical properties of the underlying memory. Thus, we do
not involve them in our analysis.

\subsection{Simulation Setup Details}
\label{memanalysis_setup}
\emph{NVMain2.0} provides an option to generate a memory trace file, which
contains detailed information for every main memory access. Using this information, we can extract the
memory address for each write access and aggregate them for each 64 byte sized
cache-line\footnote{The simulation model of \emph{gem5} assumes cache-lines to be
written to the memory entirely, hence we also use this assumption in the
analysis.}, 
which results
in a write-count distribution. 
This method is also independent from the CPU internal cache configuration, since writes to the main memory are recorded. Even if a write is caused by a logical read operation (cache preemption), this write is captured in our simulation.
We simulate an ARMv8 CPU
architecture, the DerivO3CPU implementation, and the VExpress\_GEM5\_V2 machine.
This system includes an advanced CPU with pipelining and out-of-order execution
as well as a set of controllers, which are typically found in ARM based systems
(\eg, the GIC interrupt controller, PL011 UART controller, etc.).

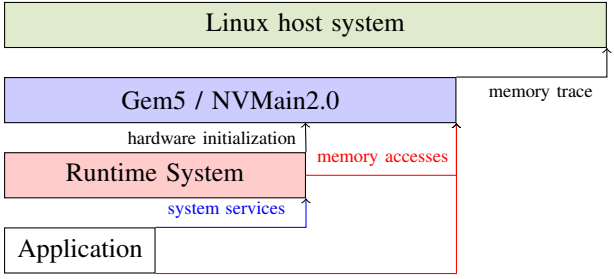
\begin{figure}
 \begin{tikzpicture}
  \node[draw, fill=TUgreen!20!white, minimum width=8cm, minimum height=.6cm] (l) at (0,3) {Linux host system};
  \node[draw, fill=blue!20!white, minimum width=6cm, minimum height=.6cm] (g) at (-1,2) {Gem5 / NVMain2.0};
  \node[draw, fill=red!20!white, minimum width=4cm, minimum height=.6cm] (r) at (-2,1) {Runtime System};
  \node[draw, fill=white!20!white, minimum width=2cm, minimum height=.6cm] (a) at (-3,0) {Application};
  
  \draw[->, draw=blue] (a.north east) -| node[above,xshift=-30] {\textcolor{blue}{\scriptsize{system services}}} (r.south east);
  
  \draw[->, draw=red] (a.south east) -| (g.south east);
  \draw[->, draw=red] (r.east) -| node[above, xshift=-28] {\textcolor{red}{\scriptsize{memory accesses}}}(g.south east);
  \draw[->, draw=black] (r.north east) -- node[left] {\textcolor{black}{\scriptsize{hardware initialization}}} (0,1.7);
  \draw[->, draw=black] (g.north east) -| node[below, xshift=-25] {\textcolor{black}{\scriptsize{memory trace}}} (l.south east);
 \end{tikzpicture}
 \caption{Overview of the simulation setup}
 \label{memanalysis_setupoverview}
\end{figure}

Two simulation modes are supported by \emph{gem5}: The systemcall-emulation and
the full system simulation. As we want to reduce the influence of the runtime
infrastructure (libraries, operating system services, etc.) on the application
as much as possible, we run \revised{bare-metal} 
full system mode simulations. 
This requires an operating system to be started in \emph{gem5}, handling the hardware
initialization and providing required services for the running application. 
We developed a small bare-metal runtime system, which takes the
place of the 
operating system in the simulation setup. Thus, we can
initialize the hardware in a flexible way with low overhead (compared to Linux kernel
modifications), and only provide the 
required operating system services.
Even if the analyzed application is directly compiled into the binary file of
the runtime system, which is started in \emph{gem5} afterwards, the runtime system can
be seen as part of the simulation environment and not as part of the
application. 
The simulation setup is illustrated in
\Cref{memanalysis_setupoverview}.

\subsection{Application Separation}
\label{memanalysis_separation}
The full system simulation mode of \emph{gem5} combined with a small, customized
runtime system in place of the operating system allows us to highly control the
hardware behavior and the memory placement. In this section, we aim to analyze
the write access behavior of an application, without interference of an
operating system, and separately analyze the memory regions of the application.
To achieve this, we apply two separation techniques:
 
\noindent\textbf{Spatial Separation}:
During the linking
process of the runtime system, the application's memory regions (\ie
\texttt{text}, \texttt{data}, \texttt{bss}, and \texttt{stack}) are placed in a
static separate memory location, which resides apart from the memory locations
of the runtime system. Thus, the memory accesses of the application target a separated
memory region, which can be analyzed separately in the recorded write-count
distribution. Furthermore, the concrete memory addresses of the memory regions
can be 
determined after the linking process, which 
allows to analyze the recorded write-count distribution separately for each
memory region. Hence, the runtime system has to establish an identity mapping (or
at least a constant, well-known mapping) from virtual memory addresses to
physical memory addresses to be able to 
determine the different
memory regions in the recorded write-count distribution. 

\noindent\textbf{Interrupt Separation}:
 The handling of interrupts is separated from the application's stack.
Usually, the operating system, respectively our runtime system, saves the
current register set on the stack when handling an interrupt. An interrupt
during the running application would cause the application's stack to be used
for the register backup, which would influence the application's write pattern
to the stack region. To overcome this, we handle interrupts on another stack
instead of the application's stack by the hardware. For ARMv8 architectures, this can
be achieved by using two different exception levels \cite{armarm}.
When taking an interrupt to a higher exception level, an ARMv8 CPU can be
configured to switch the stack pointer to a dedicated stack pointer for the
higher exception level. We run the runtime system on exception level 1 (EL1),
using a stack, allocated for the runtime system only. The application is
executed on exception level 0 (EL0) with the application's stack. Thus, whenever
an interrupt occurs during the application execution, the interrupt is handled
on EL1 on the stack of the runtime system. Accordingly, the application's stack
is not influenced by interrupts at all.


Both 
techniques allow 
to analyze the memory write-pattern of
isolated applications. Based on this, required wear-leveling actions are deduced
and proposed subsequently. In this paper, we only focus on wear-leveling for the
test applications. In a real world setup, also the runtime system / operating system
requires wear-leveling to be applied on its memory regions, because the
implementation uses the main memory similarly like the test
applications.
However, the solutions presented here can also be applied for the
runtime system, but require some additional implementation effort,  since they
are provided as a service from the runtime system itself.

\section{Aging-aware Coarse-grained Wear-Leveling}
\label{sec_coarsegrained}
\Cref{sec_appmem} points out the need for aging-aware in-memory wear-leveling, when the write-endurance is low.
If the current write behavior
cannot be tracked by the hardware and no memory trace is known for
the running application, aging-aware techniques cannot be applied. 
To overcome this issue, in this section we propose a software-only
write distribution approximation technique, 
which estimates
the memory write distribution (\ie, the write count to fixed sized memory
regions) 
using only commonly available hardware support (\ie, MMU, performance counters,
and interrupts). The write distribution approximation can be used subsequently to
enable an arbitrary aging-aware wear-leveling algorithm. However, to keep our
implementation software-only, we developed a simple aging-aware wear-leveling
algorithm, which adjusts the virtual memory mapping of the MMU to exchange the
physical location of hot (heavy written) and cold (less often written) virtual
memory pages.
Thus, the entire wear-leveling is
coarse-grained 
with a 4 kB granularity.
To omit the need of storing the aging state of the memory as a persistent object, we design our wear-leveling solution incremental. Hence, at every point in time the algorithm aims to achieve an allover write-count balance in the memory. After a reboot, for instance, the memory can be assumed to be wear-leveled and the incremental wear-leveling can be continued.
\revised{This furthermore overcomes the requirement to know the exact age of the memory at any time. Therefore, the approximation does not need to estimate absolute number, a relative representation of the write distribution is sufficient.}
At the end
of this section, we evaluate the resulting wear-leveling quality on the previously mentioned benchmark
applications.

\subsection{Write Distribution Approximation}
Several steps are required to record an approximation of the real write
distribution of an application at runtime. To achieve an equidistant sampling of write
accesses, \revised{\ie every $n^{th}$ write access is sampled}, the target of every $n^{th}$ memory write of the application 
is captured and stored in an appropriate data structure. The number $n$
determines the temporal granularity of the approximation technique, allowing a trade-off
between accuracy and introduced overhead. After capturing the write, the spatial
granularity of the data structure has to be considered as well. Storing the
estimated write count for every byte introduces a big storage overhead and leads
to imprecise results, when the temporal granularity is coarse. Instead, bytes
can be 
related to larger memory blocks and the write counts are aggregated
for every write access into these blocks. For our implementation, we aggregate the
write counts for 4 kB memory blocks, because the wear-leveling algorithm 
considers this granularity, i.e., the decision is based on memory pages.
Using an 8 byte counter for every block,
$\frac{1}{512}\cdot \text{memory-size}$ bytes are required to store the
approximated write distribution 
(\eg, 2 MB when 1GB of main memory is tracked).

The detailed flow of capturing the target of every $n^{th}$ memory write access
requires two techniques to be implemented. First, an interrupt has to be
generated after every $n^{th}$ write access, thus the runtime system can take
action. Secondly, the target of the next memory write access has to be
determined and stored in the data structure. Both implementations are stated in
detail subsequently. \revised{Although the approach by Gogte \etal allows to directly capture CPU write requests at sampled intervals \cite{227782}, their approach relies on a specialized debugging capability. Our method provides an alternative, which makes use of more widely available hardware features.}
\subsubsection{Temporal Write Distribution Sampling}
To generate an interrupt after every $n^{th}$ write access of the application,
we use the CPU internal performance counting mechanism. In ARMv8, each
performance counter can be configured to only record events triggered on EL0,
thus there is no interference of executed interrupt handlers. The
\texttt{BUS\_ACCESS\_ST} event counts the total number of store requests on the
memory bus, thus the number of write accesses of the application are recorded.
For Intel CPUs, the same behavior could be achieved by using a performance counter for writebacks of the last-level-cache. \revised{If no such performance counter is available in some system, any approximation (\eg the cycle counter), still can be considered.}
The performance counting mechanism allows to generate an interrupt when the
performance counter overflows (\ie, exceeds the value of $2^{32}-1$). To
establish interrupts on every $n^{th}$ write access, the performance counter is set
to $2^{32}-n$ during the handling of the overflow interrupt.
\subsubsection{Write Access Trapping}
As the last written memory address cannot be determined during the interrupt
handling of the performance counter overflow, a second technique is implemented
to track the target address of the the next memory write. During the handling of
the overflow interrupt, the memory access permission for the tracked memory
region is set to \texttt{READ\_ONLY}. Note that the ARMv8 architecture allows
hierarchical memory access permissions, allowing to configure memory regions of
1 GB size to \texttt{READ\_ONLY} by only modifying one page-table entry. Due to
the \texttt{READ\_ONLY} permission, the next write access causes a permission
violation trap, which is handled as an interrupt. The violation causing address
is available for the interrupt handler in a dedicated register, which then is
used to increment the corresponding counter in the write distribution
approximation\footnote{The semantics of the performance counter and of the write access trapping mechanism differ slightly. While the performance counter counts every write to the memory, including cache writebacks and other indirect memory accesses, the write access trapping only applies to CPU write operations, which require a fetch of a TLB line. However this only implies that not the target of every $n^{th}$ write is recorded,
but that sometimes the distance between two recorded writes is
$n+x$, where $x$ is a small integer.}.
During the handling of the trap, the access permissions are set back to
\texttt{READ\_WRITE}\footnote{For our runtime system implementation, memory
permissions are not used for any protection purposes. If this is the case, the
modified permissions might have to be backed up and restored later on.}.
\revised{Note that this mechanism does not strictly require a MMU, it could also be implemented with a very lightweight MPU on a microcontroller.}

\subsection{Wear-leveling Algorithm}
As mentioned before, the write distribution approximation 
enables
arbitrary aging-aware wear-leveling algorithms. When this technique is used, the
integration of the approximation system and the wear-leveling algorithm has to
be considered as well. To provide a common interface, the approximation
implementation could provide the estimated write-counts in a table inside the
runtime system's memory and a notification mechanism to trigger the
wear-leveling algorithm when a special event occurs (\eg, one estimated counter
exceeds a configured threshold). However, to reduce the overhead further, we
interleave our wear-leveling algorithm further with the approximation
implementation to reduce redundantly stored data. Our wear-leveling algorithm
uses a red-black tree \cite{Bayer1972} to maintain all managed virtual memory
pages along with their estimated age. As the estimated age is already present inside
of the tree nodes, there is no need to store these values in the approximation
implementation as well. 
\subsubsection{Management of Memory Pages}
Our wear-leveling algorithm is based on a red-black tree as the management data
structure, 
which contains all managed physical memory pages
together with their estimated cell age. Whenever a virtual memory page should be
relocated to another physical memory page, the current minimum is extracted from
the tree as the target physical page and the estimated ages are adjusted
accordingly. Regarding the overhead, the wear-leveling algorithm is only called
in this setup, when a memory page has to be relocated. Regarding the selection
policy of the wear-leveling decisions, the estimated age of all physical pages
is balanced equally over time, because every page will be the current minimum
page at a certain time when the estimated age is updated properly.

 \begin{figure}[ht!]
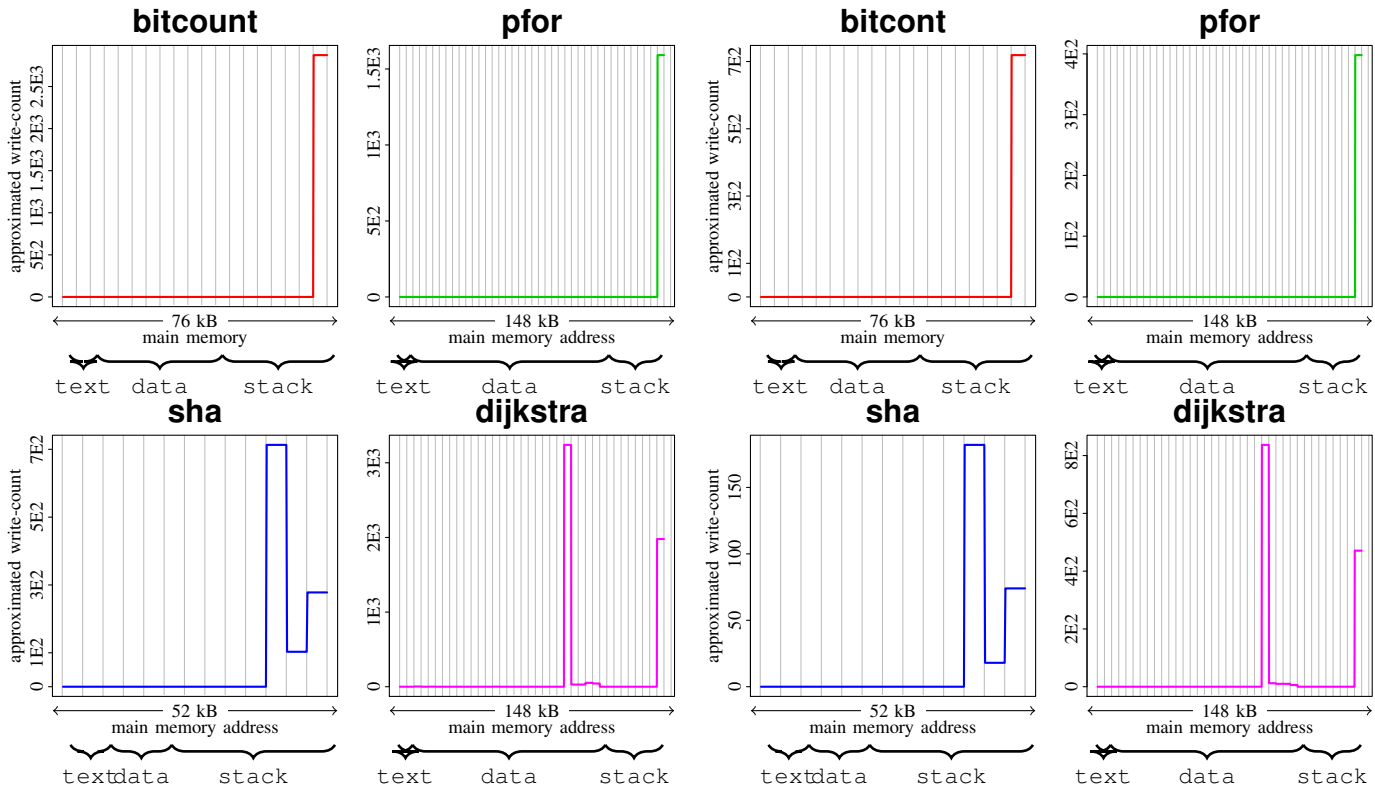

	\begin{center}
		 \begin{minipage}{.49\textwidth}
			\begin{minipage}{.49\textwidth}
				\input{figures/bitcount_approx_5000_2.tikz}
			\end{minipage}
			\begin{minipage}{.49\textwidth}
				\input{figures/pfor_approx_5000_2.tikz}
			\end{minipage}
		\end{minipage}
		
	\begin{tikzpicture}
	\newcommand{\scalefac}{1}
	\node[draw=white, rectangle, minimum width=1pt, minimum height=1pt] () at (-0.45cm,0) {};
	
	\draw [decoration={brace, amplitude=5pt}, decorate, line width=1pt] (10.5pt/\scalefac,0pt) -- node[below, yshift=-5.5pt] {\texttt{\small{text}}} (0pt/\scalefac,0pt);
	
	\draw [decoration={brace, amplitude=5pt}, decorate, line width=1pt] (57.8pt/\scalefac,0pt) -- node[below, yshift=-5pt] {\texttt{\small{data}}} (10.5pt/\scalefac,0pt);
	
	\draw [decoration={brace, amplitude=5pt}, decorate, line width=1pt] (100pt/\scalefac,0pt) -- node[below, yshift=-5pt] {\texttt{\small{stack}}} (57.8pt/\scalefac,0pt);
	
	\newcommand{\gap}{124pt}
	
	\draw [decoration={brace, amplitude=5pt}, decorate, line width=1pt] (5pt/\scalefac+\gap,0pt) -- node[below, yshift=-5.5pt] {\texttt{\small{text}}} (0pt/\scalefac+\gap,0pt);
	
	\draw [decoration={brace, amplitude=5pt}, decorate, line width=1pt] (80pt/\scalefac+\gap,0pt) -- node[below, yshift=-5pt] {\texttt{\small{data}}} (5pt/\scalefac+\gap,0pt);
	
	\draw [decoration={brace, amplitude=5pt}, decorate, line width=1pt] (100pt/\scalefac+\gap,0pt) -- node[below, yshift=-5pt] {\texttt{\small{stack}}} (80pt/\scalefac+\gap,0pt);
	\end{tikzpicture}
	
	\begin{minipage}{.49\textwidth}
			\begin{minipage}{.49\textwidth}
				\input{figures/sha_approx_5000_2.tikz}
			\end{minipage}
			\begin{minipage}{.49\textwidth}
				\input{figures/dijkstra_approx_5000_2.tikz}
			\end{minipage}
	\end{minipage}
	\begin{tikzpicture}
	\newcommand{\scalefac}{1}
	\node[draw=white, rectangle, minimum width=1pt, minimum height=1pt] () at (-0.45cm,0) {};
	
	\draw [decoration={brace, amplitude=5pt}, decorate, line width=1pt] (15.3pt/\scalefac,0pt) -- node[below, yshift=-5.5pt] {\texttt{\small{text}}} (0pt/\scalefac,0pt);
	
	\draw [decoration={brace, amplitude=5pt}, decorate, line width=1pt] (38.3pt/\scalefac,0pt) -- node[below, yshift=-5pt] {\texttt{\small{data}}} (15.3pt/\scalefac,0pt);
	
	\draw [decoration={brace, amplitude=5pt}, decorate, line width=1pt] (100pt/\scalefac,0pt) -- node[below, yshift=-5pt] {\texttt{\small{stack}}} (38.3pt/\scalefac,0pt);
	
	\newcommand{\gap}{124pt}
	
	\draw [decoration={brace, amplitude=5pt}, decorate, line width=1pt] (5.4pt/\scalefac+\gap,0pt) -- node[below, yshift=-5.5pt] {\texttt{\small{text}}} (0pt/\scalefac+\gap,0pt);
	
	\draw [decoration={brace, amplitude=5pt}, decorate, line width=1pt] (78.3pt/\scalefac+\gap,0pt) -- node[below, yshift=-5pt] {\texttt{\small{data}}} (5.4pt/\scalefac+\gap,0pt);
	
	\draw [decoration={brace, amplitude=5pt}, decorate, line width=1pt] (100pt/\scalefac+\gap,0pt) -- node[below, yshift=-5pt] {\texttt{\small{stack}}} (78.3pt/\scalefac+\gap,0pt);
	\end{tikzpicture}
		
	\end{center}
	\caption{Memory write-count approximation $n=5000$}
	\label{coarsegrained_approx_5000}
\end{figure}

Eventually, this integration of the wear-leveling algorithm and the approximation system leads to an additional configuration parameter, besides the temporal and spatial granularity of the write-count
approximation. The threshold, after which number of estimated writes a
 relocation should be performed is 
 maintained by the
 approximation system, because the wear-leveling algorithm is called from the
 approximation system in that case. This configuration parameter provides a
 trade-off between the overhead of page relocation and the frequency,
 respectively the resulting quality, of wear-leveling actions without taking
 influence on the quality of the write-count approximation.
 \subsubsection{Memory Page Relocation}
 Once the wear-leveling algorithm determined a pair of two virtual memory pages
 to swap, 
  two steps are required to perform the relocation. 
  First, the virtual
 memory mapping in the page-table has to be adjusted accordingly, such that the
 physical pages of both virtual memory pages are exchanged. A Translation
 Lookaside Buffer (TLB) maintenance operation is required afterwards to make
 sure the exchanged mapping is applied. Note that the ARMv8 virtual memory
 system allows single entries to be invalidated in the TLB, thus a total TLB
 flush is not necessary. After the new page mapping is established, the physical
 content has to be exchanged to maintain the application's view on the virtual
 memory. This is achieved by copying one page to a spare buffer, copy the second
 page to the first page, and 
 copy the buffer content to the second page.
 The size of the buffer is chosen to 4 kB for two reasons: First, copying a
 sequential memory content can be done more efficiently in most systems than
 copying single bytes or words from different regions. Second, the
 write access pattern to the buffer memory page is completely uniform and 
 thus has no negative influence on the memory lifetime if it is also handled by the wear-leveling system.
 \subsection{Evaluation}
 \label{sec_coarsegrained_eval}
 To point out how the previously presented techniques can be used to improve the
 balance of wear-levels, the write-count approximation system is evaluated
 first. 
 The four benchmark applications shown in
 \Cref{memoryanalysis_baseline_dist} are executed again with enabled write-count
 approximation. Instead of triggering the wear-leveling algorithm, the
 write-counts are simply aggregated, resulting in an analyzable distribution.
 The spatial granularity is fixed to 4 kB sized memory regions (virtual memory
 page size), while the temporal granularity is evaluated for 
 two different values. 
 For the first experiment, a sample is recorded every $n=5000^{th}$ memory
 write access, for the second experiment a sample is recorded every $n=20000^{th}$
 memory write access.
The resulting approximated write-count distributions are 
illustrated
in \Cref{coarsegrained_approx_5000} and \Cref{coarsegrained_approx_20000}.

 \begin{figure}[t]
	\begin{center}
		 \begin{minipage}{.49\textwidth}
			\begin{minipage}{.49\textwidth}
				\input{figures/bitcount_approx_20000_2.tikz}
			\end{minipage}
			\begin{minipage}{.49\textwidth}
				\input{figures/pfor_approx_20000_2.tikz}
			\end{minipage}
		\end{minipage}
		
	\begin{tikzpicture}
	\newcommand{\scalefac}{1}
	\node[draw=white, rectangle, minimum width=1pt, minimum height=1pt] () at (-0.45cm,0) {};
	
	\draw [decoration={brace, amplitude=5pt}, decorate, line width=1pt] (10.5pt/\scalefac,0pt) -- node[below, yshift=-5.5pt] {\texttt{\small{text}}} (0pt/\scalefac,0pt);
	
	\draw [decoration={brace, amplitude=5pt}, decorate, line width=1pt] (57.8pt/\scalefac,0pt) -- node[below, yshift=-5pt] {\texttt{\small{data}}} (10.5pt/\scalefac,0pt);
	
	\draw [decoration={brace, amplitude=5pt}, decorate, line width=1pt] (100pt/\scalefac,0pt) -- node[below, yshift=-5pt] {\texttt{\small{stack}}} (57.8pt/\scalefac,0pt);
	
	\newcommand{\gap}{124pt}
	
	\draw [decoration={brace, amplitude=5pt}, decorate, line width=1pt] (5pt/\scalefac+\gap,0pt) -- node[below, yshift=-5.5pt] {\texttt{\small{text}}} (0pt/\scalefac+\gap,0pt);
	
	\draw [decoration={brace, amplitude=5pt}, decorate, line width=1pt] (80pt/\scalefac+\gap,0pt) -- node[below, yshift=-5pt] {\texttt{\small{data}}} (5pt/\scalefac+\gap,0pt);
	
	\draw [decoration={brace, amplitude=5pt}, decorate, line width=1pt] (100pt/\scalefac+\gap,0pt) -- node[below, yshift=-5pt] {\texttt{\small{stack}}} (80pt/\scalefac+\gap,0pt);
	\end{tikzpicture}
	
	\begin{minipage}{.49\textwidth}
			\begin{minipage}{.49\textwidth}
				\input{figures/sha_approx_20000_2.tikz}
			\end{minipage}
			\begin{minipage}{.49\textwidth}
				\input{figures/dijkstra_approx_20000_2.tikz}
			\end{minipage}
	\end{minipage}
	\begin{tikzpicture}
	\newcommand{\scalefac}{1}
	\node[draw=white, rectangle, minimum width=1pt, minimum height=1pt] () at (-0.45cm,0) {};
	
	\draw [decoration={brace, amplitude=5pt}, decorate, line width=1pt] (15.3pt/\scalefac,0pt) -- node[below, yshift=-5.5pt] {\texttt{\small{text}}} (0pt/\scalefac,0pt);
	
	\draw [decoration={brace, amplitude=5pt}, decorate, line width=1pt] (38.3pt/\scalefac,0pt) -- node[below, yshift=-5pt] {\texttt{\small{data}}} (15.3pt/\scalefac,0pt);
	
	\draw [decoration={brace, amplitude=5pt}, decorate, line width=1pt] (100pt/\scalefac,0pt) -- node[below, yshift=-5pt] {\texttt{\small{stack}}} (38.3pt/\scalefac,0pt);
	
	\newcommand{\gap}{124pt}
	
	\draw [decoration={brace, amplitude=5pt}, decorate, line width=1pt] (5.4pt/\scalefac+\gap,0pt) -- node[below, yshift=-5.5pt] {\texttt{\small{text}}} (0pt/\scalefac+\gap,0pt);
	
	\draw [decoration={brace, amplitude=5pt}, decorate, line width=1pt] (78.3pt/\scalefac+\gap,0pt) -- node[below, yshift=-5pt] {\texttt{\small{data}}} (5.4pt/\scalefac+\gap,0pt);
	
	\draw [decoration={brace, amplitude=5pt}, decorate, line width=1pt] (100pt/\scalefac+\gap,0pt) -- node[below, yshift=-5pt] {\texttt{\small{stack}}} (78.3pt/\scalefac+\gap,0pt);
	\end{tikzpicture}
		
	\end{center}
	\caption{Memory write-count approximation $n=20000$}
	\label{coarsegrained_approx_20000}
\end{figure}

\subsubsection{Write-Count Approximation Evaluation}
The characteristic of the real write-count distribution
(compared to \Cref{memoryanalysis_baseline_dist}) is reflected properly in both
experiments. The main peaks inside the distribution are 
shown regarding
their height compared to the rest of the distribution. The variation of the
temporal granularity can be observed due to the different scaling of the y axes.
\revised{Since our approach performs incremental wear-leveling, the total memory lifetime is not considered. Hence, the absolute scaling of the write approximation does not matter.}
However, the reduction of the temporal granularity does not influence the
preciseness of the approximation in this setup, because still enough samples are
recorded, even for $n=20000$. If the application executes relative short or the
temporal granularity is configured too coarse, not enough samples might be
available to reflect the characteristic of the distribution properly. 
This trade-off should be taken into account when considering the temporal
granularity.
\begin{table}[h]
    \centering
    \begin{tabular}{c r r r r}
         &\textbf{bitcount}&\textbf{pfor}&\textbf{sha}&\textbf{dijkstra}\\
         \hline
         $n=5000$&5.72\%&11.50\%&4.94\%&7.20\%\\
         $n=20000$&1.50\%&3.24\%&1.77\%&1.89\%\\
         \hline
    \end{tabular}
    \caption{CPU overhead for the write-count approximation}
    \label{coarsegrained_approx_overhead}
\end{table}

When choosing a temporal granularity, the introduced overhead
should be also considered. To evaluate the overhead, the necessary additional CPU cycles are calculated as a percentage of the baseline execution, without write-count
approximation.
\Cref{coarsegrained_approx_overhead} lists the calculated CPU overhead of  both
experiments. 
The relative overhead is similar
for all benchmarks, because the approximation system 
reacts  relative to the total write count, respectively the execution time. 


\subsubsection{Full Wear-Leveling Evaluation}
To 
determine if the estimation is precise enough to enable aging-aware
wear-leveling, the approximation and wear-leveling algorithm is plugged together and evaluated again.
The red-black tree based wear-leveling algorithm is activated and triggered from
the approximation system. The spatial granularity remains at 4 kB while the
temporal granularity of the approximation again is chosen as $n=5000$ and
$n=20000$. A remapping of a page is requested, whenever the write-count
estimation exceeds the value of $4$ (for $n=5000$) or the value of $1$
($n=20000$). This leads to mostly the same total number of page relocations in
both experiments. Thus they can be compared regarding the quality of the write
count approximation. 
\begin{figure}[t!]
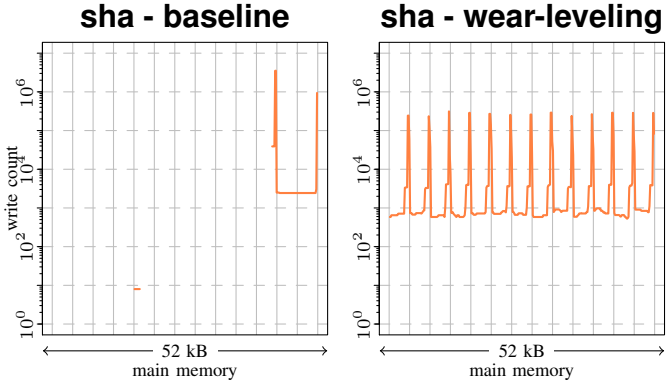

	\begin{center}
		 \begin{minipage}{.49\textwidth}
			\begin{minipage}{.49\textwidth}
				\input{figures/sha_baseline_log.tikz}
			\end{minipage}
			\begin{minipage}{.49\textwidth}
				\input{figures/sha_balanced_5000_4_log.tikz}
			\end{minipage}
		\end{minipage}
	\end{center}
	\caption{Coarse-grained full Wear-Leveling Result For sha $n=5000$}
	\label{coarsegrained_result_5000}
\end{figure}

\begin{figure}[t!]
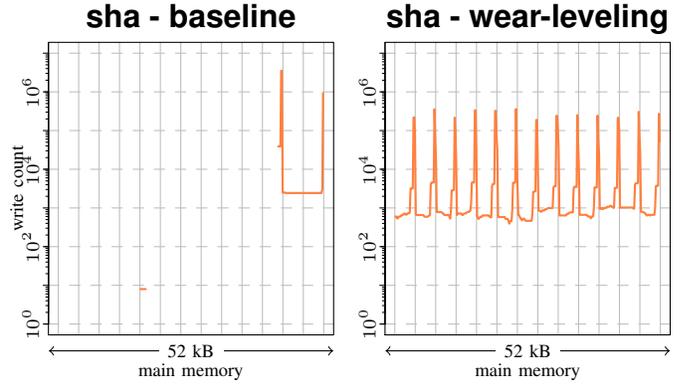

	\begin{center}
		 \begin{minipage}{.49\textwidth}
			\begin{minipage}{.49\textwidth}
				\input{figures/sha_baseline_log.tikz}
			\end{minipage}
			\begin{minipage}{.49\textwidth}
				\input{figures/sha_balanced_20000_1_log.tikz}
			\end{minipage}
		\end{minipage}
	\end{center}
	\caption{Coarse-grained full wear-leveling result for sha $n=20000$}
	\label{coarsegrained_result_20000}
\end{figure}

\Cref{coarsegrained_result_5000} and \Cref{coarsegrained_result_20000} show the
resulting write distribution 
of our simulation under 
coarse-grained wear-leveling 
for the \textbf{sha} benchmark. 
The results from the other benchmarks are only presented by their calculated
 improvement later due to space limitation. Note that due to the logarithmic
 scale of the y axes memory bytes with a write-count of 0 are not displayed. 
The estimated write-count distribution is precise enough to perform
aging-aware relocations and 
balance the wear-levels across the
target memory region. 
\subsubsection{Memory Lifetime Improvement}
\label{coarsegrained_li_sec}
Considering the gained improvement of the memory lifetime
requires some assumptions. First, the system is considered dead once the first
memory cell is worn out. Thus, the maximum write count to the memory determines
the memory lifetime. Assuming that the target of each write access could be
shuffled through the memory arbitrarily, the theoretical best memory lifetime
could be achieved when every memory cell is written equally often, thus the mean
write count would be applied to each cell. Combining both considerations,
\Cref{coarsegrained_eq_ae} calculates the achieved endurance ($AE$), which is
the fraction of the ideal memory lifetime, which is achieved by the analyzed
execution. A value of $1$ means that the experiment already achieves the maximum memory lifetime, while a value of, for instance, $0.5$ means that the memory lifetime could be doubled in the ideal case.
\begin{equation}
    AE=\frac{\text{mean\_write\_count}}{\text{max\_write\_count}}
    \label{coarsegrained_eq_ae}
\end{equation}
Comparing the achieved endurance of an execution with enabled wear-leveling to
the baseline  without any wear-leveling leads to an endurance improvement
($EI$),  which can be determined according to \Cref{coarsegrained_eq_ei}. The maximum endurance improvement thus depends on the achieved endurance of the baseline.
\begin{equation}
    EI=\frac{AE_\text{analyzed}}{AE_\text{baseline}}
    \label{coarsegrained_eq_ei}
\end{equation}
The endurance improvement 
describes how 
many additional write accesses can be 
performed before the memory wears out while using the analyzed wear-leveling 
technique, compared to the baseline, but does not give any insight if the 
application profits from the additional writes. For instance, an $EI$ of 2 means that the application can perform twice as many writes compared to the situation without wear-leveling. If the wear-leveling causes $100\%$ overhead, all the additional writes would be consumed by the wear-leveling and no real benefit would be achieved. Therefore, the 
introduced write overhead $WO$ (as a percentage of the total number of  writes
of the baseline execution) has to be considered as well to determine 
the lifetime improvement ($LI$) according to \Cref{coarsegrained_eq_li}. A $LI$ value of, for instance, $2$ implies that the application can perform twice as much writes, respectively can run twice as long, regardless of introduced overhead and writes for the wear-leveling.
\begin{equation}
    LI=\frac{EI}{WO+1}
    \label{coarsegrained_eq_li}
\end{equation}
Similarly, the achieved endurance can be related to the write overhead, which leads to the normalized endurance ($NE$).
\begin{equation}
 NE=\frac{AE}{WO+1}
\end{equation}
The write overhead is determined in this evaluation on the simulation 
results by comparing the total number of memory writes for each benchmark 
execution with the corresponding baseline. For the four benchmark applications, 
the achieved endurance, the write overhead and the lifetime improvement is 
calculated for both wear-leveling experiments and the results are collected  in
\Cref{coarsegrained_result_tab}. 
\begin{table}[h!]
    \centering
    \begin{tabular}{c r r r r r}
         &&$AE$&$WO$&$NE$&\textcolor{black}{$LI$}\\
         \hline
         $n=5000$&\textbf{bitcount}&0.016&5.10\%&0.015&\textcolor{black}{18.90}\\
         &\textbf{pfor}&0.043&5.10\%&0.041&\textcolor{black}{40.01}\\
         &\textbf{sha}&0.022&5.05\%&0.021&\textcolor{black}{11.20}\\
         &\textbf{dijkstra}&0.022&5.10\%&0.021&\textcolor{black}{28.65}\\
         $n=20000$&\textbf{bitcount}&0.016&5.11\%&0.015&\textcolor{black}{18.93}\\
         &\textbf{pfor}&0.044&5.12\%&0.042&\textcolor{black}{40.06}\\
         &\textbf{sha}&0.019&5.10\%&0.018&\textcolor{black}{9.72}\\
         &\textbf{dijkstra}&0.022&5.11\%&0.021&\textcolor{black}{28.26}\\
         \hline
    \end{tabular}
    \caption{Lifetime improvement ($LI$) for coarse-grained wear-leveling}
    \label{coarsegrained_result_tab}
\end{table}

We observe the following properties. 
First, the memory write
overhead is mostly independent from the configuration of the approximation
system, because the approximation in general does not cause many additional
memory writes. Second, the lifetime improvement depends on the total amount of
memory which is used for the wear-leveling, since the write pattern of the
application is anyway mostly targeting a single memory page. If this page can be
remapped to colder pages, the improvement 
is higher. 
Third, although
the lifetime is improved by a considerable factor, the achieved endurance
remains at mostly $\approx 4\%$ of the ideal lifetime in all benchmarks. This 
stems from the high non-uniformity within memory pages, which is caused by
the applications. As memory pages are only relocated to other 4~kB aligned
memory pages, the non-uniformity within pages is not resolved by the wear-leveling
system.

To summarize this section, aging-aware wear-leveling on the coarse-granularity
of 4 kB sized memory pages 
performs reasonably in a software-only manner due to the statistical
write-count  approximation. Nevertheless, a coarse-grained wear-leveling
technique alone 
is not sufficient to achieve an equal balance of 
the wear-levels allover the memory due to the high non-uniformity within memory pages.

\section{Fine-grained Stack Wear-Leveling}
\label{sec_finegrained}
To overcome the problem of intra page non-uniformity, solutions in literature
are extended with a finer grained wear-leveling technique, resolving the
non-uniformity in the scope of coarse-grained memory regions, which are targeted
by the coarse-grained technique subsequently \cite{startgap,
Zhou:2009:DEE:1555754.1555759}. 
To the best of our knowledge, all the fine-grained
extensions are \revised{either} realized in hardware by remapping single bytes or group of bytes
with an additional abstraction \revised{or by functional data remapping \cite{jacobvitz2014coset}, which requires at least compiler support}. In this section, we propose a software-only
fine-grained extension to the coarse-grained wear-leveling system
(\Cref{sec_coarsegrained}), which resolves non-uniform write accesses in the
memory pages of the stack region. These pages are targeted by the coarse-grained
wear-leveling system subsequently and are remapped to other physical pages.

Since
all fine-grained wear-leveling extensions are hardware based,
we most likely cannot propose a generic fine-grained wear-leveling approach 
based on
commonly available hardware. Instead, we propose a specialized technique,
which only targets the stack region of C / C++ compiled applications.
The concept to target the stack with a specialized wear-leveling system in a
software-based manner is also 
considered by Li \etal \cite{stackalloc}. The
basic idea is to allocate every stack frame for a new function call on the heap
through an aging-aware memory allocator. This approach features two major
disadvantages: First, the wear-leveling quality relies on the application to
perform enough and fine-grained function calls to apply sufficient wear-leveling
actions. Second, the amount of required stack memory might not be known in
advance\footnote{C99 allows dynamic sized local arrays \cite{gccvarar}. However,
this could also be achieved in assembly.}, which leads to a certain
fragmentation and to worse wear-leveling results. Due to these disadvantages, we  in
contrast
relocate the entire stack memory without the application's cooperation.

As the stack is used by the compiled code relative to the stack pointer
(\texttt{sp})\footnote{Depending on the application logic, concrete pointer
values may be also calculated and stored in variables. These pointer are also considered when the memory location of the stack is changed.}, the
application can be instructed to use another memory location as the stack by
adjusting the \texttt{sp}. As the stack anyway 
is the main cause
for non-uniform write accesses 
(see \Cref{sec_coarsegrained_eval}), we focus our
fine-grained wear-leveling extension on relocating the stack to other memory
locations and thus resolve the non-uniform write access pattern inside the
stack.

\subsection{Circular Stack Relocation}
\label{finegrained_sec_circular}
To evenly distribute the write accesses to the stack, we
move the stack region in a circular manner through the memory.
In essence, the physical memory content is relocated
with a fixed offset into one direction always with an overflow semantics at the
end of the memory.
For the \emph{Start-gap} approach, this can be achieved by a
corresponding remapping function, because an additional abstraction layer
maintains the logical view on the memory. 
The runtime system allocates a memory region of the size of multiple memory pages for the application's stack.
The stack is
relocated from time to time by setting the \texttt{sp} further by an offset and
copying the old stack content to the according new location. The logical view of
the application always expects free memory bytes left (negative offset) of the \texttt{sp} and the
already created stack content directly right (positive offset) of the~\texttt{sp}. As long as the
stack only is relocated into one direction, this view can be maintained easily.
A wraparound at the end of the reserved memory region cannot be achieved
trivially when the stack should be relocated by the same offset in each step, since the stack content cannot be split.
Thus, we install a mechanism, called shadow stack, which aids to implement the
wraparound at the end of the reserved memory region.
\subsubsection{Shadow Stack} The basic concept of the shadow stack is to allow
one part of the stack to maintain at the end of the reserved memory region,
while the rest of the stack already is wrapped around to the beginning. At any
point in time, the entire stack content must be accessible by addressing memory
contents right of the \texttt{sp} (with a positive offset). Furthermore, at any
point in time the same amount of free memory should be available left of the
\texttt{sp} (with a negative offset). Only by maintaining these two properties,
the application can continue the execution at any time.
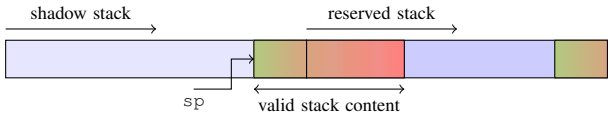
\begin{figure}
    \centering
    \begin{tikzpicture}
        
        \node[draw, fill=blue!20!white, minimum width=4cm, minimum height=0.5cm] () at (2,0) {};
        \node[draw, fill=blue!10!white, minimum width=4cm, minimum height=0.5cm] () at (-2,0) {};
        
        \draw[->] (0,0.4) -- node[above] {\scriptsize{reserved stack}} (2,0.4);
        \draw[->] (-4,0.4) -- node[above] {\scriptsize{shadow stack}} (-2,0.4);
        
        \node[draw, shading=axis, left color=TUgreen!50!red!50!white, right color=red!50!white, minimum width=1.3cm, minimum height=0.5cm] () at (0.65,0) {};
        \node[draw, shading=axis, left color=TUgreen!100!red!50!white, right color=TUgreen!50!red!50!white, minimum width=0.7cm, minimum height=0.5cm] () at (-0.35,0) {};
        
        \node[draw, shading=axis, left color=TUgreen!100!red!50!white, right color=TUgreen!50!red!50!white, minimum width=0.7cm, minimum height=0.5cm] () at (3.65,0) {};
        
        \draw[<->] (-0.7,-0.4) -- node[below] {\scriptsize{valid stack content}}(1.3,-0.4);
        
        \draw[->] (-1.5,-0.4) node[below] {\scriptsize{\texttt{sp}}} -| (-1,-0.4) |- (-0.7,0);
        
    \end{tikzpicture}
    \vspace{-0.3cm}
    \caption{Shadow stack}
    \vspace{-0.5cm}
    \label{finegrained_shadowstack}
\end{figure}

The setup of the shadow stack is 
 illustrated in
\Cref{finegrained_shadowstack}. Technically, the real stack is present as a
consecutive virtual memory region, which is shown in the right half of
\Cref{finegrained_shadowstack}. For the shadow stack, the same amount of virtual
memory space left of the real stack is allocated and is mapped to exactly the
same physical memory pages like the real stack. Thus, given an arbitrary virtual
address $\mathcal{A}$ of the real stack, the same physical content is accessed
at the virtual address $\mathcal{S}(\mathcal{A})=\mathcal{A}-\text{stacksize}$. This also
implies that setting the \texttt{sp} from some virtual address $\mathcal{S}(\mathcal{A})$
inside the shadow stack to the corresponding real stack address
$\mathcal{A}$ does not change the application's perspective on the stack at
all. Using this mechanism, the stack relocation is implemented in two steps.
First, the stack is moved down the memory periodically. At any time, the
application can access the same amount of memory left of the \texttt{sp},
because the writes can target the shadow stack. Once the currently used stack
(including all valid stack content) is entirely moved to the shadow stack, the
\texttt{sp} is set back to the corresponding real stack address. As mentioned before, the virtual memory at the new location of the \texttt{sp} contains exactly the same content as at the old location.
Hence, the application's perspective is maintained and the entire stack is wrapped around back to the real stack (right half).
Repeating these two steps regularly, the stack is relocated in a circular manner with
the same offset in each relocation step.
\subsubsection{Combination with Coarse-grained Wear-Leveling}
\label{finegraned_sec_combinatiom}
As stated before, the fine-grained wear-leveling is designed as an extension to
the previously presented coarse-grained wear-leveling system
(\Cref{sec_coarsegrained}). Both systems can work together 
nearly out of the box. Since the stack relocation only operates in the virtual memory space, a stack relocation can  only be  interrupted by the remapping of the page to another physical memory page.
Nevertheless, when remapping hot
and cold pages, the coarse-grained wear-leveling system has to be aware of the
special shadow stack configuration and has to maintain it during remapping.
Furthermore, the statistical write-count approximation has to aggregate the
captured write accesses from the shadow stack and from the real stack to the
same physical page. Eventually, we set up a frequent stack relocation by using the same performance counter overflow interrupt mechanism like the coarse-grained wear-leveling system. This ensures that stack relocations are triggered after a certain number of writes to the memory. Additionally, 
the overhead can be reduced by combining the interrupt mechanism and only using
one interrupt service routine (ISR).

\subsection{Address Consistency}
The concept of moving the stack in a circular manner
(\Cref{finegrained_sec_circular}) is based on the \texttt{sp} relative access of
the stack region by C / C++ compiled applications. 
However, the \texttt{sp} relative access is not the only way to access memory
contents within the stack memory. Sometimes, the application requires to create pointers
to variables inside the stack to pass it to subsequent function calls or to
store the pointer in a central variable. Furthermore, pointers to variables on
the stack may also be moved out of the stack to some global or heap data
structures. 
During a relocation of the stack, the memory address of
the variables on the stack changes, while the content of the pointers stays
unchanged. This leads to invalid pointers and to a wrong behavior of the
application. To overcome this problem, we equip the fine-grained relocation
system with two pointer adjustment mechanisms, which maintain the correctness of
pointer contents over stack relocations.
\subsubsection{In-memory Pointer Adjustment}
First, an in-memory pointer adjustment technique targets pointers to stack
contents, which are stored inside the stack itself. This is the usual
 case
when pointers to local variables are passed to subsequent function calls or
positions inside local arrays need to be remembered. For the relocation of
the stack, the entire valid stack content has to be copied to the new memory
location anyway, resulting in every memory word from the current valid stack is
loaded to the CPU and stored back to the memory. During this process,
 the memory word is checked, 
 and a pointer to  stack variable is adjusted by the
relocation offset. 
To identify a memory word as a pointer into the
stack, a strong constraint needs to be put to the memory usage of the
application. As the memory word is just seen as a 8 byte number by the
relocation routine, the application has to make sure to not use any logic
variable content, which has the same number like a pointer value into the stack
would have. We ensure this by allocating the virtual memory pages of the stack
at a memory location bigger than 4 GB and allow the application to use 64 bit
aligned data types with the 32 lower bits set only.
\subsubsection{Smart-Pointer Adjustment}
As the previous technique only targets pointers, which are stored inside the
stack, pointers which are stored in global or heap data structures still are
corrupted after a stack relocation. To solve this problem, the fine-grained
wear-leveling system ships with a smart-pointer implementation, which checks the
current relocation of the stack during dereferencing. The internally stored raw
pointer is adjusted properly and dereferenced. The smart-pointer implementation
only allows to hand out copied variables, but not the internal raw pointer.
Whenever the application aims to move a pointer out of the stack, it has to use
the smart-pointer implementation instead of a raw pointer.


To summarize, maintaining the consistency of pointers during stack relocations
puts strong constraints on the application and blows up in-memory data
structures. Nevertheless, the constraints can be achieved by reimplementing
applications accordingly and this enables software-only fine-grained in-memory
wear-leveling.

\subsection{Evaluation}
\begin{figure}[t!]
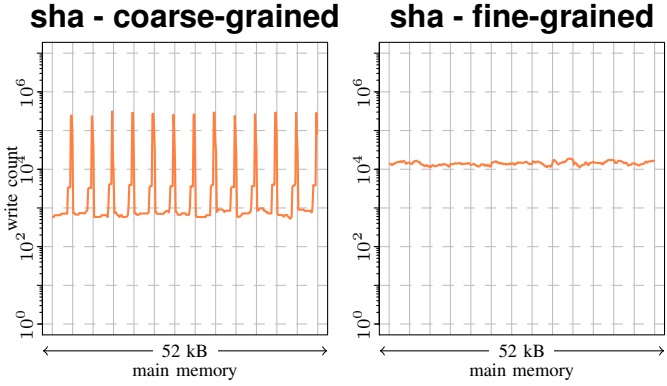

	\begin{center}
		 \begin{minipage}{.49\textwidth}
			\begin{minipage}{.49\textwidth}
				\input{figures/sha_balanced_5000_4_log_2.tikz}
			\end{minipage}
			\begin{minipage}{.49\textwidth}
				\input{figures/sha_stack_1000_64_log.tikz}
			\end{minipage}
		\end{minipage}
	\end{center}
	\caption{Fine-grained wear-leveling result for sha (page relocation every $t=64^{th}$ stack relocation)}
	\vspace{-0.5cm}
	\label{finegrained_result_64}
\end{figure}
The technical details 
of the combined implementation of the fine-grained
stack relocation technique and the coarse-grained aging-aware wear-leveling
system are explained in \Cref{finegraned_sec_combinatiom}. The movement of the
stack by an offset of 64 bytes\footnote{In our simulation setup 64 byte
cache-lines are assumed to be written entirely. A finer movement than 64 byte
has no further effect on the wear-leveling result in this case.} is triggered
periodically from the performance counter overflow mechanism. In this evaluation
the performance counter overflow is configured to trigger after every
$n=1000^{th}$ memory write access, thus the stack is relocated every $1000^{th}$
memory write. Accordingly, the write-count approximation works on the same
temporal granularity. The coarse-grained wear-leveling system is triggered
whenever a page exceeds an approximated write-count of $t=64$ and thus in mean
on every $64^{th}$ stack relocation. Considering the relocation offset of $64$
bytes, a coarse-grained page relocation is triggered whenever the stack is
relocated by $4096$ bytes, which is the size of one memory page. A second
experiment is executed with the trigger for the coarse-grained wear-leveling
system set to $t=32$. This increases the total number of page relocations at the
cost of higher memory overhead. Furthermore, in this scenario page relocations
are performed when the stack only passed half of a memory page size, thus the
internal non-uniformity is higher.

\begin{figure}[t!]
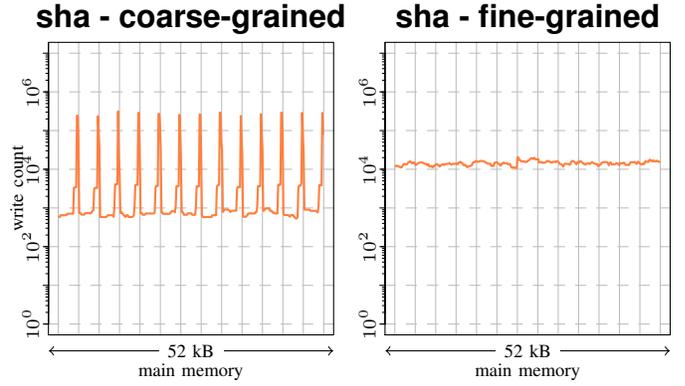

	\begin{center}
        \begin{minipage}{.49\textwidth}
			\begin{minipage}{.49\textwidth}
				\input{figures/sha_balanced_5000_4_log_2.tikz}
			\end{minipage}
			\begin{minipage}{.49\textwidth}
				\input{figures/sha_stack_1000_32_log.tikz}
			\end{minipage}
		\end{minipage}
	\end{center}
	\caption{fine-grained wear-leveling result for sha (page relocation every $t=32^{nd}$ stack relocation)}
	\vspace{-0.5cm}
	\label{finegrained_result_32}
\end{figure}

\Cref{finegrained_result_64} and \Cref{finegrained_result_32} show the resulting
memory write-count distribution for the \textbf{sha} benchmark, compared to the
coarse-grained wear-leveling system only (\Cref{coarsegrained_result_5000}) for
both benchmark configurations. The 
results show  that the non-uniformity
within virtual memory pages can be resolved by the fine-grained stack
wear-leveling technique and thus the allover write pattern to the main memory is
more uniform. Even though the total number of page relocations is higher in the
second experiment (\Cref{finegrained_result_32}), the results from the first
experiment are slightly better due to the fact that a page relocation is only
performed, when the stack is moved by an offset of an entire memory page.
\subsubsection{Memory Lifetime Improvement}
To finalize the evaluation, the improvement of the memory lifetime can be
calculated in the same way like in \Cref{coarsegrained_li_sec}. The
according results are collected in \Cref{finegrained_result_tab}.
\begin{table}[h!]
    \centering
    \begin{tabular}{c r r r r r}
         &&$AE$&$WO$&$NE$&\textcolor{black}{$LI$}\\
         \hline
         $t=64$&\textbf{bitcount}&0.788&0.47\%&0.784&\textcolor{black}{953.52}\\
         &\textbf{pfor}&0.698&9.17\%&0.639&\textcolor{black}{614.45}\\
         &\textbf{sha}&0.746&111.59\%&0.353&\textcolor{black}{187.55}\\
         &\textbf{dijkstra}&0.018&2.90\%&0.017&\textcolor{black}{23.64}\\
         $t=32$&\textbf{bitcount}&0.592&0.79\%&0.587&\textcolor{black}{713.98}\\
         &\textbf{pfor}&0.462&10.78\%&0.417&\textcolor{black}{400.96}\\
         &\textbf{sha}&0.693&112.91\%&0.328&\textcolor{black}{173.09}\\
         &\textbf{dijkstra}&0.020&4.50\%&0.019&\textcolor{black}{25.87}\\
         \hline
    \end{tabular}
    \caption{Lifetime improvement (LI) for fine-grained wear-leveling}
    \vspace{-0.5cm}
    \label{finegrained_result_tab}
\end{table}
First of all, it can be observed that the write overhead $WO$ has a high
variation for the different benchmarks. This is caused by the different way of
stack usage by each benchmark. The \textbf{sha} application for instance uses a
big part of the stack memory and thus has a very high write overhead. 
The total write distribution of the application in the end determines the
lifetime improvement $LI$. The \textbf{dijkstra} application for instance also
faces a high non-uniform memory usage within the bss segment, which is
not resolved by our fine-grained wear-leveling technique. Thus, the results for
\textbf{dijkstra} are relative bad.

In conclusion, the memory lifetime can be improved significantly, if the intra page non-uniformity can be resolved by the
fine-grained stack wear-leveling, \eg, $\approx900$ times for the \textbf{bitcount} application.
Note that the memory lifetime improvement strongly depends on the
available memory size. In this evaluation, only the minimal required amount of
memory for each benchmark is considered. If a system offers additional spare
memory, the memory lifetime can be further improved. The improvement is
determined mostly by the resulting uniformity of the memory access distribution
($AE$) and the write overhead.

\subsubsection{Comparison to the Literature}
Several techniques for in-memory wear-leveling for NVM have been proposed over
the last years. In this section we compare our evaluation results with following
related techniques: \emph{Start-gap} was proposed by Qureshi \etal
\cite{startgap} and relocates the entire memory space in a circular manner on the granularity of 256
byte cache-lines through special hardware. To resolve non-uniformity within
cache-lines, a finer-grained address space randomization is introduced. Khouzani
\etal \cite{AghaeiKhouzani:2014:PPL:2627369.2627667} proposed a wear-leveling
scheme, which hooks into the page allocation process of the operating system.
Due to knowledge about the current write-count and the write characteristic to
each memory region, wear-leveling actions are decided and performed. Chen \etal
\cite{Chen:2012:APW:2228360.2228439} proposed a similar scheme with advanced
management data structures to make the wear-leveling algorithm more efficient.
This approach only operates on the coarse granularity of virtual memory pages.

As a metric, we adopted the term normalized endurance ($NE$) from the
\emph{Start-gap} approach, which is our achieved endurance value related to the
memory write overhead.
As a concrete lifetime or a relative improvement always highly depends on the
considered benchmark and the memory size, we use the normalized endurance as a
fraction of the possible ideal memory usage, respectively the memory lifetime.
Unfortunately only a few works consider the possible ideal lifetime in their
evaluation. 
The previously mentioned works \cite{startgap, AghaeiKhouzani:2014:PPL:2627369.2627667, Chen:2012:APW:2228360.2228439} 
all report to achieve
almost the ideal memory lifetime in the best case (\ie, in the range of $\approx 87\%$ to $\approx 98\%$). 
Our best result achieves $78.43\%$ of the
ideal memory lifetime.


%


As our system requires no additional hardware and can be tuned regarding the
write-overhead, it enables a trade-off for the design-process of a hardware
platform. The necessary costs for the required hardware support for in-memory
wear-leveling can be replaced by the slightly worse wear-leveling quality and a
possibly bigger runtime overhead

\section{Outlook on Further Fine-Grained Extensions}
The final evaluation results in \Cref{finegrained_result_tab} show that the all-over wear-leveling quality can be good, if the non-uniformity of write accesses within memory pages can be resolved. However, not only the stack has to be targeted by a fine-grained specific extension, but also the {data}/{bss} and, if it exists, the heap segment. For instance, the \textbf{dijkstra} application has a highly non-uniform memory usage inside the {bss} segment leading to a bad performance. The {text} segment requires no special wear-leveling, because all accesses are read-only by definition. While specific wear-leveling for the {heap} has been targeted in form of aging-aware memory allocations in the literature \cite{stackalloc, DBLP:conf/asplos/CoburnCAGGJS11}, the {data}/{bss}  segment requires another special technique. For future work, we propose to relocate elements of the {data}/{bss} segment by using the feature of dynamic linked code. If the application is not statically linked, the addresses or an access offset for the {data}/{bss} segment is determined and set while the application is loaded. During a maintenance phase, \ie, an interrupt, the {text} segment could be re-loaded with relocated addresses of the {data}/{bss} segment and thus these segments can be relocated.
This could achieve
a circular movement, similar to the movement for the stack, for the {data}/{bss} segment. 


\section{Conclusion}
\label{sec_conclusion}
Recently, several in-memory wear-leveling techniques have been proposed to
tackle a major disadvantage, namely the lower write endurance, of NVM
technologies, which might replace classic DRAM in the near future. Advanced,
aging-aware wear-leveling techniques rely on  hardware-provided age information,
such as a write-count per cell / byte / domain, to achieve good wear-leveling
results. As the necessary hardware support is not available in common or
commercial off-the-shelf (COTS) hardware, it introduces additional costs. The
hardware at least requires additional chip-space, but also might be very complex
to build to meet a certain clock-speed and granularity.

To overcome the need for this hardware and offer the possibility to use the
chip-space for other features, this paper introduced a software-only,
aging-aware wear-leveling system, which only makes use of widely available
hardware features. The final evaluations show that we are able to achieve up
to $78.43\%$ of the theoretically ideal possible memory lifetime with our
wear-leveling system without any additional hardware costs. During the design
process of a system, it might be totally reasonable to only achieve roughly
$80\%$ of the possible memory lifetime (\eg 8 instead of 10 years), but to equip
the system with advanced hardware controllers to improve energy
consumption, for instance.

As we believe it is important to offer the possibility for such software-only
in-memory wear-leveling, we release all our sources, including benchmark applications and wear-leveling implementations: \url{https://github.com/tu-dortmund-ls12-rt/NVMSimulator}.

\section*{Acknowledgement}
This paper is supported in parts by the German Research Foundation (DFG) Project OneMemory (Project number 405422836).



\newpage
\bibliographystyle{IEEEtranS}
\bibliography{references}

\end{document}